\documentclass[preprint*]{JHEP3} 

\JHEPspecialurl{http://jhep.sissa.it/JOURNAL/JHEP3.tar.gz}

\usepackage{epsfig,multicol}

\newcommand\fverb{\setbox\pippobox=\hbox\bgroup\verb}
\newcommand\fverbdo{\egroup\medskip\noindent%
			\fbox{\unhbox\pippobox}\ }
\newcommand\fverbit{\egroup\item[\fbox{\unhbox\pippobox}]}
\newbox\pippobox

\title{Wess-Zumino model with exact supersymmetry on the lattice}

\author{Marisa Bonini and Alessandra Feo\\
	Dipartimento di Fisica, Universit\`a di Parma and INFN Gruppo Collegato di Parma,
       Parco Area delle Scienze, 7/A. 43100 Parma, Italy \\
	E-mail: \email{bonini@pr.infn.it}, \email{feo@fis.unipr.it}}

\preprint{UPRF-2004-04}

\abstract{A lattice formulation of the four dimensional Wess-Zumino model that uses Ginsparg-Wilson fermions 
and keeps exact supersymmetry is presented. The supersymmetry transformation that leaves invariant the action at finite lattice 
spacing is determined by performing an iterative procedure in the coupling constant. 
The closure of the algebra, generated by this transformation is also showed. }


\begin{document} 

\newcommand{\beeq}{\begin{equation}}
\newcommand{\eneq}{\end{equation}}
\newcommand{\beeqa}{\begin{eqnarray}}
\newcommand{\eneqa}{\end{eqnarray}}

\section{Introduction}
Non-perturbative studies of supersymmetric theories turn out to have remarkably rich properties 
which are of great physical interest. For this reason, much effort has been dedicated to formulating
a lattice version of supersymmetric theories. See for example 
\cite{dondi}-\cite{montvay} and \cite{kaplan,feo} for recent reviews.
While much is known analytically, the hope is that the lattice would provide further information 
and confirm the existing analytical calculations.

The major obstacle in formulating a supersymmetric theory on the lattice arises from the fact that 
the supersymmetry algebra is actually an extension of the Poincar\'e algebra, which is explicitly 
broken by the lattice.
Indeed, in an interacting theory, translation invariance is broken since the Leibniz rule is not valid for
lattice derivatives \cite{dondi}. Ordinary Poincar\'e algebra is also broken by the lattice but
the hypercubic crystal symmetry forbids relevant operators which could spoil the Poincar\'e symmetry 
in the continuum limit.
In the case of the super Poincar\'e algebra, the lattice crystal
group is not enough to guarantee the absence of supersymmetry violating operators.
Without exact lattice supersymmetry one might hope to construct non-supersymmetric lattice theories with 
a supersymmetric continuum limit. 
This is the case of the Wilson fermion approach for the $N=1$ supersymmetric 
Yang-Mills theory where the only operator which violates the $N=1$ supersymmetry is a fermion mass term.
By tunning the fermion mass to the supersymmetric limit one recovers supersymmetry in the continuum 
limit \cite{montvay,curci}.
Alternatively, using domain wall fermions \cite{dwf} or overlap fermions \cite{overlap}, this fine tunning 
is not required. 
Recently, a new lattice construction for models with extended supersymmetry 
has been proposed \cite{kaplan2}. In this case, the lattice preserves some supersymmetries which are enough to 
reduce or eliminate the need for fine tunning (see also \cite{catterall2}).

In the past the lattice Wess-Zumino model has been perturbatively studied using Wilson fermions
\cite{bartels,golterman} and adding to the action a Wilson term also for the
scalar fields. In the continuum limit this results in a cancellation
of divergences between fermion and scalar fields. However, scalar and
fermion renormalization wave functions in general do not coincide, due
to finite contributions, thus in order to restore supersymmetry in the
continuum limit a fine tuning of the various coupling of the lattice
action is needed \cite{bartels}.  For the two dimensional case this
problem is not present \cite{golterman}, at least in perturbation
theory, where the continuum supersymmetric Ward identities are
recovered in the limit of vanishing lattice spacing without a fine
tuning~\footnote{ Non-perturbative effects may produce supersymmetry
breaking at finite volume\cite{catterallkaramov}.}.

More recently, a lattice Wess-Zumino model has been defined in
Refs.~\cite{fujikawa2,fujikawa} using a general Ginsparg-Wilson
operator. In this case the supersymmetric continuum limit is
recovered without a fine-tuning also in four dimensions \cite{fujikawa}.
Moreover, this formulation allows to consider Yukawa interactions
which are invariant under lattice chiral transformation
\cite{luscher}, thus it appears to be suitable for chiral theories
and, in particular, for supersymmetric gauge theories.
 
In this paper we consider the four dimensional lattice Wess-Zumino
model introduced in Refs.~\cite{fujikawa2,fujikawa} and show that it
is actually possible to formulate the theory in such a way that the
full action is invariant under a lattice superymmetry transformation
at a fixed lattice spacing.  The action and the transformation are
written in terms of the Ginsparg-Wilson operator and reduce to their
continuum expression in the naive continuum limit $a\to 0$.  The
lattice supersymmetry transformation is non-linear in the scalar
fields and depends on the parameters $m$ and $g$ entering in the
superpotential.  
We also show that the lattice supersymmetry transformation close the algebra, which is a necessary ingredient to
guarantee the request of supersymmetry.
We believe that the existence of this exact symmetry
is responsible for the restoration of supersymmetry in the continuum
limit, which has been explicitly verified in perturbation theory in
the case of the scalar and fermion two-point functions \cite{fujikawa}.

The paper is organized as follows. In Sec.~\ref{sec2} we introduce the Ginsparg-Wilson fermion operator and 
formulate the lattice Wess-Zumino action. In Sec.~\ref{sec3} we show how to build up a lattice supersymmetry transformation 
that is an exact symmetry of this model. 
In Sec.~\ref{sec4} the closure of the algebra, crucial step to be satisfied in order to impose supersymmetry,
is shown. Discussions and outlook are summarized in Sec.~\ref{sec5}.
In Appendices~A,~B and ~C, some details of the calculations are presented.

\section{The Wess-Zumino model}
\label{sec2}
The Ginsparg-Wilson relation \cite{ginsparg}
\beeq
\gamma_5 D + D \gamma_5 = a D \gamma_5 D  
\label{gw}
\eneq
implies a continuum symmetry of the fermion action which may be regarded as a lattice form of the chiral 
symmetry \cite{luscher}. As a matter of fact, the fermion lagrangian with a Yukawa interaction 
\beeq
{\cal L} = \bar \psi D \psi + g \bar \psi (P_{+} \phi \hat P_{+} +  P_{-} \phi^\dagger \hat P_{-}) \psi \, ,
\label{yukawa}
\eneq
where 
\beeq
P_\pm = \frac{1}{2} (1 \pm \gamma_5) \, , \qquad \qquad   \hat P_\pm  = \frac{1}{2} (1 \pm \hat \gamma_5)
\eneq
are the lattice chiral projection operators and $\hat \gamma_5 = \gamma_5 (1 - a D)$, is invariant under 
the lattice chiral transformation 
\beeq
\delta \psi = i \varepsilon \hat \gamma_5 \psi \, , \qquad \qquad \delta \bar \psi = i \bar \psi \gamma_5 \varepsilon 
 \, , \qquad \qquad  \delta \phi = -2 i \varepsilon \phi \, .
\eneq

By writing $\psi$  in terms of two Majorana fermions 
\beeq
\psi = \chi + i \eta \, ,
\eneq
it can be seen that the interaction term in Eq.~(\ref{yukawa}) couples the two Majorana fermions and 
therefore there is a conflict between lattice chiral symmetry and the Majorana condition \cite{pvn,fujikawa2}. This is due to the fact 
that the projection operators $\hat P_{\pm}$ depend on $D$. 
Moreover, it has been observed that by making the following field redefinition 
\beeq
\psi' = (1 - \frac{a}{2} D) \psi \, , \qquad \qquad \bar \psi' = \bar \psi \, ,
\eneq
the Yukawa interaction becomes
\beeq
 g \bar \psi' (P_{+} \phi P_{+} +  P_{-} \phi^\dagger  P_{-}) \psi' 
\eneq
and the two Majorana components of $\psi'$ decouple. 
Taking advantage of this property, one can define the four dimensional Wess-Zumino on the lattice with Majorana fermions 
\cite{fujikawa2}. 

We start with a lagrangian defined in terms of the Ginsparg-Wilson fermions on the $d=4$ euclidean lattice.
Our analysis is valid for all operators which satisfy Eq.~(\ref{gw}), however, in the following we will use the 
particularly simple solution given by \cite{neuberger}
\beeq
D = \frac{1}{a} \bigg( 1 - \frac{X}{\sqrt{X^\dagger X}} \bigg) \, , \qquad \qquad X = 1 - a D_w \, ,
\label{D}
\eneq
\noindent 
where
\beeq
D_w = \frac{1}{2} \gamma_\mu ( \nabla^\star_\mu + \nabla_\mu ) - \frac{a}{2} \nabla^\star_\mu \nabla_\mu 
\label{Dw}
\eneq
and 
\beeqa
\nabla_\mu \phi(x) &=& \frac{1}{a}(\phi(x + a \hat \mu) - \phi(x)) \nonumber \\  
\nabla_\mu^\star \phi(x) &=& \frac{1}{a}(\phi(x) - \phi(x - a \hat \mu))   
\eneqa
are the forward and backward lattice derivatives, respectively.
\noindent
Substituting Eq.~(\ref{Dw}) in Eq.~(\ref{D}) we find convenient to isolate in $D$ the part containing the gamma matrices 
and write 
\beeq
D = D_1 + D_2 
\eneq
where 
\beeq
D_1 = \frac{1}{a} \bigg( 1 - \frac{1 + \frac{a^2}{2} \nabla^\star_\mu \nabla_\mu}{\sqrt{X^\dagger X}} \bigg) \, , \qquad \qquad 
D_2 = \frac{1}{2} \gamma_\mu \frac{\nabla^{\star}_\mu + \nabla_\mu}{\sqrt{X^\dagger X}} \equiv  \gamma_\mu D_{2 \mu} \, .
\eneq
In terms of $D_1$ and $D_2$ the Ginsparg-Wilson relation (\ref{gw}) becomes
\beeq
D_1^2 - D_2^2 = \frac{2}{a} D_1 \, .
\label{gw1}
\eneq

The action of the 4-dimensional Wess-Zumino model on the lattice has been introduced in Refs.~\cite{fujikawa2,fujikawa} 
and can be re-written using the above notation as 
\beeqa
S_{WZ} &=&  \sum_x \bigg \{ \frac{1}{2} \bar \chi (1 - \frac{a}{2} D_1)^{-1} D_2 \chi - 
\frac{2}{a} \phi^{\dagger}D_1 \phi + F^{\dagger} (1 - \frac{a}{2} D_1)^{-1} F + \frac{1}{2} m \bar \chi \chi \nonumber \\
&& + m (F \phi + (F \phi)^\dagger) + g \bar \chi (P_{+} \phi P_{+} + P_{-} \phi^{\dagger} P_{-}) \chi + 
g (F \phi^2 + (F \phi^2)^\dagger) \bigg\} \, ,
\label{wz}
\eneqa
\noindent
where $\phi$ and $F$ are scalar fields and $\chi$ is a Majorana fermion which satisfies the Majorana condition 
\beeq
\bar \chi = \chi^T C
\eneq
\noindent
and $C$ is the charge conjugation matrix which satisfies 
\beeq
C^T = -C \, , \qquad \qquad C C^\dagger = 1 \, .
\eneq
Moreover, our conventions are 
\beeqa
&& C \gamma_\mu C^{-1} = - (\gamma_\mu)^T  \nonumber \\
&& C \gamma_5 C^{-1} = (\gamma_5)^T  \, .
\label{c}
\eneqa
\noindent

It is easy to see that in the continuum limit, $a \to 0$, Eq.~(\ref{wz}) reduces to the continuum Wess-Zumino
action
\beeqa
S & =& \int \bigg\{ \frac{1}{2} \bar \chi (\not \partial + m) \chi + \phi^\dagger \partial^2 \phi +  F^{\dagger} F + 
m (F \phi + (F \phi)^\dagger) \nonumber \\ 
&& + g \bar \chi (P_{+} \phi P_{+} + P_{-} \phi^{\dagger} P_{-}) \chi  + g (F \phi^2 + (F \phi^2)^\dagger) \bigg\} \, .
\eneqa

\section{The supersymmetric transformation}
\label{sec3}
If one defines the real components by
\beeq
\phi \to \frac{1}{\sqrt{2}} (A + i B) \, , \, \, \, \, \, \, F \to \frac{1}{\sqrt{2}} (F - i G) 
\eneq
the Wess-Zumino action (\ref{wz}) can be written as 
\beeq
S_{WZ} = S_0 + S_{int} \, , 
\label{wz2}
\eneq
with 
\beeqa
&& S_{0} = \sum_x \bigg \{ \frac{1}{2} \bar \chi (1 - \frac{a}{2} D_1)^{-1} D_2 \chi - 
\frac{1}{a} ( A D_1 A + B D_1 B)  \nonumber \\ 
&& \phantom{S_{0} = \sum_x \bigg \{ }+ \frac{1}{2} F (1 - \frac{a}{2} D_1)^{-1} F + \frac{1}{2} G (1 - \frac{a}{2} D_1)^{-1} G 
\bigg\} \label{wz0} \, , \\[12 pt]
&& S_{int} = \sum_x \bigg\{ \frac{1}{2} m \bar \chi \chi + m (F A + G B) 
+ \frac{1}{\sqrt{2}} g \bar \chi (A + i \gamma_5 B) \chi \nonumber \\ 
&& \phantom{S_{int} = \sum_x \bigg \{ }+ \frac{1}{\sqrt{2}} g \big[ F (A^2 - B^2) + 2 G (A B) \big] \bigg\} \, .
\label{wzint}
\eneqa

The free part of the action, $S_0$, is invariant under the lattice supersymmetry transformation
\beeqa
&& \delta A = \bar \varepsilon \chi = \bar \chi \varepsilon  \nonumber \\
&& \delta B = -i \bar \varepsilon \gamma_5 \chi = -i \bar \chi \gamma_5 \varepsilon \nonumber \\
&& \delta \chi = - D_2 (A - i \gamma_5 B) \varepsilon - (F - i \gamma_5 G) \varepsilon 
\nonumber \\ 
&& \delta F = \bar \varepsilon D_2 \chi  \nonumber \\
&& \delta G = i \bar \varepsilon  D_2 \gamma_5 \chi \, .
\label{susytransf}
\eneqa
\noindent
In fact, the variation of $S_0$ under the this transformation is 
\beeqa
&& \delta S_0 = \nonumber \\
&& = \sum_x \Big\{ \bar \chi (1 - \frac{a}{2} D_1)^{-1} D_2 \Big[-D_2 (A - i \gamma_5 B) \varepsilon
- (F - i \gamma_5 G) \varepsilon \Big] - \frac{2}{a} \bar \chi \varepsilon D_1 A \nonumber \\
&& \phantom{= \sum_x \Big\{ }+ \frac{2 i}{a} \bar \chi \gamma_5 \varepsilon D_1 B + 
(\bar \varepsilon D_2\chi) (1 - \frac{a}{2} D_1)^{-1}  F 
+ i (\bar \varepsilon D_2 \gamma_5 \chi) (1 - \frac{a}{2} D_1)^{-1} G \Big\} \, . \nonumber 
\eneqa
By using (\ref{c}) and integrating by part~\footnote{For instance, for any scalar function ${\cal F}$ one has
${\cal F} \bar \varepsilon D_2 \chi = \bar \chi D_2 {\cal F} \varepsilon$.  }, this variation becomes
\beeqa
&& \sum_x  \Big\{ -\bar \chi \varepsilon \bigg[ (1 - \frac{a}{2} D_1)^{-1} D_2^2 + \frac{2}{a} D_1 \bigg] A 
+ i \bar \chi \gamma_5 \varepsilon \bigg[ (1 - \frac{a}{2} D_1)^{-1} D_2^2 + \frac{2}{a} D_1 \bigg] B \nonumber \\
&& - \bar \chi (1 - \frac{a}{2} D_1)^{-1} D_2 (F - i \gamma_5 G) \varepsilon
+ \bar \chi D_2 \varepsilon (1 - \frac{a}{2} D_1)^{-1}  F 
+ i \bar \chi D_2 \gamma_5 \varepsilon (1 - \frac{a}{2} D_1)^{-1} G \Big\} \nonumber \\
&& = 0 \, , \nonumber 
\eneqa
where we have used the Ginsparg-Wilson relation (\ref{gw1}), which implies 
\beeq
(1 - \frac{a}{2} D_1)^{-1} D_2^2 = - \frac{2}{a} D_1 \, . \nonumber 
\eneq
As discussed in \cite{fujikawa}, the variation of $S_{int}$ under ~(\ref{susytransf}) does not vanish because of the 
failure of the Leibniz rule at finite lattice spacing \cite{dondi}.

In order to discuss the symmetry properties of the lattice Wess-Zumino model one possibility is to modify the 
action by adding irrelevant terms which make invariant the full action.
Alternatively, one can modify the supersymmetry transformation of Eq.~(\ref{susytransf}) in such a way that the action 
(\ref{wz2}) has an exact symmetry for $a$ different from zero
\footnote{A similar attempt has been proposed by Golterman and Petcher, \cite{golterman}, for the 2-dimensional 
Wess-Zumino model.}. Since the transformation (\ref{susytransf}) leaves invariant the free part of the action, 
this modification must vanish for $g=0$. Therefore, we introduce the following transformation
\beeqa
&& \delta A = \bar \varepsilon \chi = \bar \chi \varepsilon  \nonumber \\
&& \delta B = -i \bar \varepsilon \gamma_5 \chi = -i \bar \chi \gamma_5 \varepsilon \nonumber \\
&& \delta \chi = - D_2 (A - i \gamma_5 B) \varepsilon - (F - i \gamma_5 G) \varepsilon + 
g R \varepsilon \nonumber \\ 
&& \delta F = \bar \varepsilon D_2 \chi \nonumber \\
&& \delta G = i \bar \varepsilon D_2 \gamma_5 \chi 
\label{complete}
\eneqa
where $R$ is a function to be determined by requiring that the variation of the action vanishes. We make the 
assumption that $R$ depends on the scalar and auxiliary fields and their derivatives and not on $\chi$.

The variation of the Wess-Zumino action under the transformation (\ref{complete}) is 
\beeqa
\delta S_{WZ} &=& \sum_x \Big\{ g \bar \chi (1 - \frac{a}{2} D_1)^{-1} D_2 R \varepsilon - m \bar \chi \Big[ D_2 (A - i \gamma_5 B) \varepsilon 
+ (F - i \gamma_5 G) \varepsilon - g R \varepsilon \Big] \nonumber \\
&& + m (A \bar \varepsilon D_2 \chi + F \bar \chi \varepsilon +
i B \bar \varepsilon D_2 \gamma_5 \chi - i G \bar \chi \gamma_5 \varepsilon) 
+ \frac{g}{\sqrt{2}} \bar \chi (\bar \varepsilon \chi + \gamma_5 (\bar \varepsilon \gamma_5 \chi)) \chi \nonumber \\
&& - \sqrt{2} g \bar \chi (A + i \gamma_5 B) \Big[D_2 (A - i \gamma_5 B) \varepsilon
+ (F - i \gamma_5 G) \varepsilon - g R \varepsilon \Big] \nonumber \\ 
&& + \frac{g}{\sqrt{2}}  \Big[ (A^2 - B^2) \bar \varepsilon D_2 \chi +
2 F A \bar \chi \varepsilon + 2 i F B \bar \chi \gamma_5 \varepsilon \nonumber \\
&& + 2 i A B \bar \varepsilon D_2 \gamma_5 \chi + 2 G B \bar \chi \varepsilon 
- 2 i G A (\bar \chi \gamma_5 \varepsilon) \Big] \Big\} \nonumber \, .
\eneqa
By using the Fierz identity, terms with four fermions cancel as in the continuum. Moreover, $g$ independent terms
cancel out after an integration by part, and one is left with 
\beeqa
\delta S_{WZ} &=& \sum_x \Big\{ g \bar \chi \Big[(1 - \frac{a}{2} D_1)^{-1} D_2 R + m R \Big] \varepsilon 
- \frac{g}{\sqrt{2}} \Big[ 2 \bar \chi (A + i \gamma_5 B) D_2 (A - i \gamma_5 B) \varepsilon \nonumber \\
&& \phantom{\sum_x \Big\{ } - \bar \chi D_2 (A - i \gamma_5 B)^2 \varepsilon \Big] 
+ \sqrt{2} g^2 \bar \chi (A + i \gamma_5 B) R \varepsilon \Big\} \, .
\label{intvar}
\eneqa
\noindent 
The function $R$ is determined by imposing the vanishing of $\delta S_{WZ}$. By expanding $R$ in powers of $g$ 
\beeq
R = R^{(1)} + g R^{(2)} + \cdots
\label{expansion}
\eneq
and imposing the symmetry condition order by order in perturbation theory, we find 
\beeq
R^{(1)} = ((1 - \frac{a}{2} D_1)^{-1} D_2 + m )^{-1} \Delta L 
\label{r1}
\eneq
with 
\beeqa
&& \Delta L \equiv \frac{1}{\sqrt{2}} ( 2 (A + i \gamma_5 B) D_2 (A - i \gamma_5 B) - D_2 (A - i \gamma_5 B)^2 ) \nonumber \\
&& \phantom{\Delta L} = \frac{1}{\sqrt{2}} \Big\{2 (A D_2 A - B D_2 B) - D_2 (A^2 - B^2) \nonumber \\ 
&& \phantom{\Delta L = \frac{1}{\sqrt{2}} \Big\{ } + 2 i \gamma_5 \Big[(A D_2 B + B D_2 A) - D_2 (A B)\Big] \Big\} \, .
\eneqa
To order $g^2$ one has
\beeq
R^{(2)} = -\sqrt{2} ((1 - \frac{a}{2} D_1)^{-1} D_2 + m)^{-1} (A + i \gamma_5 B) ((1 - \frac{a}{2} D_1)^{-1} D_2 + m)^{-1} 
\Delta L \, ,
\label{r2}
\eneq
and for $n \geq 2$
\beeq
R^{(n)} = - \sqrt{2} ((1 - \frac{a}{2} D_1)^{-1} D_2  + m)^{-1} (A + i \gamma_5 B) R^{(n-1)}  \, .
\label{rn}
\eneq
By inserting these results in Eq.~(\ref{expansion}), the function $R$ to be used in Eq.~(\ref{complete})
is therefore, the formal solution of 
\beeq
\big[(1 - \frac{a}{2} D_1)^{-1} D_2  + m  + \sqrt{2} g (A + i \gamma_5 B) \big] R = \Delta L \, . 
\label{dl}
\eneq
\noindent
Notice that, from the perturbative expressions (\ref{r1}) and (\ref{rn}) one realizes that $R \to 0$ for 
$a \to 0$, since $\Delta L$ vanishes in this limit. Indeed, $\Delta L$ is different from zero because
of the breaking of the Leibniz rule for a finite lattice spacing.

\section{The algebra}
\label{sec4}
We now study the algebra associated to the lattice supersymmetry transformation (\ref{complete})
introduced in the previous section. 
In particular, carrying out the commutator of two supersymmetries we must find a transformation which is still a 
symmetry of the Wess-Zumino action, i.e. the transformations of the fields form a closed algebra, order
by order in $g$. In this section, we explicitly check this fact up to order $g^1$, even though the calculation can
be generalized to any order.

Two supersymmetry transformations on the scalar field $A$ give
\beeqa
&& \delta_1 \delta_2 A = \delta_1 (\bar \varepsilon_2 \chi) \nonumber \\
&& \phantom{\delta_1 \delta_2 A }= - \bar \varepsilon_2 \big[ D_2 (A - i \gamma_5 B) \varepsilon_1 +
(F - i \gamma_5 G) \varepsilon_1 - g R \varepsilon_1 \big] \nonumber 
\eneqa
and their commutator yields
\beeq
[\delta_2,\delta_1] A = -2 \bar \varepsilon_1 D_2 \varepsilon_2 A + 
g (\bar \varepsilon_1 R \varepsilon_2 - \bar \varepsilon_2 R \varepsilon_1) \, . 
\label{com}
\eneq
The order $g^1$ of the second term on the r.h.s. of (\ref{com}) reads 
\beeqa
&& g (\bar \varepsilon_1 R^{(1)} \varepsilon_2 -  \bar \varepsilon_2 R^{(1)} \varepsilon_1) = \nonumber \\
&& \sqrt{2} g \bar \varepsilon_2 \frac{m(1 - \frac{a}{2} D_1)}{m^2(1 - \frac{a}{2} D_1) + \frac{2}{a} D_1} 
\Big[D_2 (A^2 - B^2) - 2 (A D_2 A - B D_2 B) \Big] \varepsilon_1 
\eneqa
where we used (\ref{r1}). Then, the commutator of two supersymmetries on the scalar field $A$ is 
\beeqa
&& \! \! \! \! \! \! \! \! \! [\delta_2,\delta_1] A = -2 \bar \varepsilon_1 \gamma_\mu \varepsilon_2 \Big\{  D_{2 \mu} A \nonumber \\ 
&& \! \! \! \! \! \! \! \! \! \phantom{[\delta_2,\delta_1] A = }+ \frac{g}{\sqrt{2}}  \frac{m(1 - \frac{a}{2} D_1)}{m^2(1 - \frac{a}{2} D_1) + 
\frac{2}{a} D_1} 
\Big[D_{2 \mu}  (A^2 - B^2) - 2 (A D_{2 \mu} A - B D_{2 \mu} B) \Big] \Big\} \, .
\label{delta12A}
\eneqa
Similarly, the commutators of two supersymmetries on the other fields, up to terms of order $g^1$, 
are (see Appendix~A for some details)
\beeqa
&&  \! \! \! \! \! \! \! \! \! [\delta_2,\delta_1] B = -2 \bar \varepsilon_1 \gamma_\mu \varepsilon_2 \Big\{  D_{2 \mu} B \nonumber \\ 
&&  \! \! \! \! \! \! \! \! \! \phantom{[\delta_2,\delta_1] B = } + \sqrt{2} g \frac{m(1 - \frac{a}{2} D_1)}{m^2(1 - \frac{a}{2} D_1) + 
\frac{2}{a} D_1} 
\Big[D_{2 \mu} (A B) - (A D_{2 \mu} B + B D_{2 \mu} A) \Big] \Big\} \, ,
\label{delta12B} \\
&& \! \! \! \! \! \! \! \! \! [\delta_2,\delta_1] F = -2 \bar \varepsilon_1 \gamma_\mu \varepsilon_2 \Big\{  D_{2 \mu} F \nonumber \\ 
&& \! \! \! \! \! \! \! \! \! \phantom{[\delta_2,\delta_1] F = }- \frac{g}{\sqrt{2}}  \frac{D_2^2}{m^2(1 - \frac{a}{2} D_1) + \frac{2}{a} D_1} 
\Big[D_{2 \mu}  (A^2 - B^2) - 2 (A D_{2 \mu} A - B D_{2 \mu} B) \Big] \Big\} \, , 
\label{delta12F} \\
&&  \! \! \! \! \! \! \! \! \! [\delta_2,\delta_1] G = -2 \bar \varepsilon_1 \gamma_\mu \varepsilon_2 \Big\{  D_{2 \mu} G \nonumber \\ 
&&  \! \! \! \! \! \! \! \! \! \phantom{[\delta_2,\delta_1] G = } - \sqrt{2} g \frac{D_2^2}{m^2(1 - \frac{a}{2} D_1) + \frac{2}{a} D_1} 
\Big[D_{2 \mu} (A B) - (A D_{2 \mu} B + B D_{2 \mu} A) \Big] \Big\} 
\label{delta12G} 
\eneqa
and 
\beeqa
&&  \! \! \! \! \! \! \! \! \! [\delta_2,\delta_1] \chi = -2 \bar \varepsilon_1 \gamma_\mu \varepsilon_2 \Big\{ D_{2 \mu} \chi \nonumber \\
&&  \! \! \! \! \! \! \! \! \! \phantom{[\delta_2,\delta_1] \chi = } -  
\frac{g}{\sqrt{2}} \frac{m(1 - \frac{a}{2} D_1) - D_2}{m^2(1 - \frac{a}{2} D_1) + 
\frac{2}{a} D_1}
\bigg( D_2 (A - i \gamma_5 B) \gamma_\mu \chi  + (A + i \gamma_5 B) D_2 \gamma_\mu \chi \nonumber \\ 
&& \phantom{[\delta_2,\delta_1] \chi = -  \frac{1}{\sqrt{2}} g \frac{m(1 - \frac{a}{2} D_1) - D_2}{m^2(1 - \frac{a}{2} D_1) + 
\frac{2}{a} D_1} } - D_2 \big[ (A - i \gamma_5 B) \gamma_\mu \chi \big] \bigg) \bigg\}  \, .
\label{delta12chi}
\eneqa
Therefore, the general expression of these commutators is 
\beeq
[\delta_1 , \delta_2] \Phi = \alpha^\mu P^\Phi_\mu(\Phi) \, , \qquad \qquad \Phi = (A,B,F,G,\chi) \, ,
\eneq
where $\alpha^\mu = -2 \bar \varepsilon_2 \gamma^\mu \varepsilon_2$ and $P^\Phi_\mu(\Phi)$ are polynomials in 
$\Phi$ defined as 
\beeq
P^\Phi_\mu(\Phi) = D_{2 \mu}\Phi + O(g) 
\label{deltaalpha}
\eneq
where the order $g^1$ contributions can be read in (\ref{delta12A})-(\ref{delta12chi}). 
We have verified that the closure works, i.e. the action (\ref{wz}) is invariant under the transformation 
\beeq
\Phi \to \Phi + \alpha^\mu P^\Phi_\mu(\Phi) 
\label{deltaalpha1}
\eneq
up to terms of order $g^1$. This calculation is sketched in Appendix~B. 
Notice that, in the continuum limit $D_{2 \mu} \to \partial_\mu $  and the transformation (\ref{deltaalpha1}) 
reduces to 
\beeq
\Phi \to \Phi + \alpha^\mu \partial_\mu \Phi  
\eneq
since terms in (\ref{delta12A})-(\ref{delta12chi}) of order $g^1$ vanish due to the restoration 
of the Leibniz rule as $a \to 0$.

Higher orders in $g$ of the transformation (\ref{deltaalpha1}) can be determined by using the expression 
for $R^{(n)}$ given in (\ref{rn}). 
The proof of the invariance of the Wess-Zumino action under the transformation (\ref{complete}) at any order 
in $g$ can be similarly performed. Indeed, the closure of the algebra at any order in $g$ should hold since the supersymmetry 
transformation (\ref{complete}) is an exact symmetry of the lattice Wess-Zumino action.

\section{Conclusions}
\label{sec5}
In this paper, we have presented a lattice formulation of the four dimensional Wess-Zumino model 
with an exact supersymmetry using Ginsparg-Wilson fermions. We have shown that it is actually possible to 
formulate the theory in such a way that the full action is invariant under a lattice supersymmetry
transformation at a fixed lattice spacing.  
This supersymmetry transformation introduces a function $R$ which is non-linear in the scalar fields 
and depends on the parameters $m$ and $g$ entering in the interaction part of the action. 
The action and the transformation, which have been written 
in terms of the Ginsparg-Wilson fermions, reduce to their continuum expression in the limit 
$a \to 0$. We have also shown that the lattice supersymmetry transformations close the algebra, 
as it is required by the supersymmetry. 
While the present work is confined to the proof to the order $g^1$, concerning the closure of the 
algebra, there are no obstructions to extending this procedure to higher orders in $g$. 

The study of the Ward identities associated to the exact lattice
supersymmetry we have introduced can be done by generalizing the
analysis performed by Golterman and Petcher in \cite{golterman} for
the two dimensional Wess-Zumino model. In the Appendix~C, we have calculated a simple Ward identity 
up to order $O(g)$ and verified that it is satisfied.
We believe that the lattice
supersymmetry we have introduced automatically leads to a restoration of the continuum
supersymmetry without additional fine tuning.  Explicit
results on the two point functions \cite{fujikawa} lend support to this
idea and we are currently investigating on this issue in more detail.

Obviously one the most important question is whether these ideas may
be extended to supersymmetric gauge theories where we expect that the 
Ginsparg-Wilson relation will play an important role in the construction
of a lattice supersymmetry. However, there is an
important difference between gauge theories and the Wess-Zumino model.
The free and the interaction terms of a lattice gauge action are both
contained in the plaquette term and therefore it is not obvious how to
perform the iterative construction of the lattice supersymmetry transformation
in this case.

\vspace{1cm}\noindent{\large\bf Acknowledgements} \\
We are grateful to Giuseppe~Marchesini and Mario~Pernici for enlightening and very useful discussions.

\section*{Appendix A}
In this Appendix we show some details concerning the calculation of the commutator $[\delta_2,\delta_1] \chi $. 
Applying two supersymmetry transformations on $\chi$ one has 
\beeq
\hspace{-1.4cm} \delta_1\delta_2 \chi = 
\hspace{-1.4cm} \phantom{\delta_1\delta_2 \chi }
= - D_2 \big[ \bar \varepsilon_1 \chi - \gamma_5 (\bar \varepsilon \gamma_5 \chi) \big] \varepsilon_2
- \big[ (\bar \varepsilon_1 D_2 \chi) + \gamma_5 (\bar \varepsilon_1 D_2 \gamma_5 \chi) \big] \varepsilon_2 +
g (\delta_1 R) \varepsilon_2 \, ,
\label{12chi}
\eneq
where $ (\delta_1 R) $ is the supersymmetry variation of the function $R$.
A similar expression is obtained for $\delta_2\delta_1\chi $ with $\varepsilon_2 \leftrightarrow \varepsilon_1 $.
Terms to the order $g^0$ can be treated as in the continuum case by using a Fierz rearrangement as well as 
the relations in Eq.~(\ref{c}). For instance,  
taking the first term in Eq.~(\ref{12chi}) and the corresponding one with $\varepsilon_2 \leftrightarrow \varepsilon_1 $
one has 
\beeq
-\bar \varepsilon_1 D_{2 \mu} \chi (\gamma_\mu \varepsilon_2)_\alpha +\bar \varepsilon_2 D_{2 \mu} \chi 
(\gamma_\mu \varepsilon_1)_\alpha =
\frac{1}{2} (\bar \varepsilon_1 \gamma_\mu \varepsilon_2) (D_2 \gamma_\mu \chi)_\alpha - 
\frac{1}{4}  (\bar \varepsilon_1 \gamma_{\mu \nu} \varepsilon_2) (D_2 \gamma_{\mu \nu} \chi)_\alpha \, .
\eneq
Using a similar rearrangement for the remaining terms order $g^0$, the commutator of two supersymmetry transformations on $\chi$ 
is 
\beeq
[\delta_2,\delta_1] \chi_\alpha = -2 \bar \varepsilon_1 \gamma_\mu \varepsilon_2  D_{2 \mu} \chi_\alpha + O(g) \, . 
\eneq
The contribution to this commutator to the order $g^1$ is
\beeqa
&& (\delta_1 R^{(1)} \varepsilon_2 - \delta_2 R^{(1)} \varepsilon_1) =  \frac{\delta R^{(1)}}{\delta A} \varepsilon_2 
(\bar \varepsilon_1 \chi) 
+ \frac{\delta R^{(1)}}{\delta B} \varepsilon_2 (\bar \varepsilon_1 \gamma_5 \chi) - (\varepsilon_1 \leftrightarrow \varepsilon_2 ) 
\nonumber \\
&& \phantom{(\delta_1 R^{(1)} \varepsilon_2 - \delta_2 R^{(1)} \varepsilon_1)} = 
\sum_R (\bar \varepsilon_1 \gamma_R \varepsilon_2) \bigg( \frac{\delta R^{(1)}}{\delta A} \gamma_R \chi + 
\frac{\delta R^{(1)}}{\delta B} \gamma_R \gamma_5 \chi \bigg) - (\varepsilon_1 \leftrightarrow \varepsilon_2 )
\eneqa
where the sum is over the 16 independent $4 \times 4$ matrices.
By using Eq.~(\ref{c}) only the terms with $\gamma_R = \big\{ \gamma_\mu, \gamma_{\mu \nu} \big\}$ survive, moreover, using 
the explicit form for $R^{(1)}$ one finds Eq.~(\ref{delta12chi}).

\section*{Appendix B}
The prove of the invariance of the action under the transformation (\ref{deltaalpha}) to the order $g^0$ is immediate.
In this Appendix, we explicitly calculate the variation of the fermionic part of the action (\ref{wz2}) to the order $g^1$ 
(the remaining part, containing only scalar fields, can be similarly treated).
\beeqa
&& \delta \sum_x \bigg \{ \frac{1}{2} \bar \chi (1 - \frac{a}{2} D_1)^{-1} D_2 \chi +  \frac{1}{2} m \bar \chi \chi + 
\frac{1}{\sqrt{2}} g \bar \chi (A + i \gamma_5 B) \chi \bigg\} \nonumber \\
&& = -\frac{g}{\sqrt{2}} \sum_x \bigg\{ \bar \chi \bigg( (1 - \frac{a}{2} D_1)^{-1} D_2 + m \bigg)
\frac{m(1 - \frac{a}{2} D_1) - D_2}{m^2(1 - \frac{a}{2} D_1) + \frac{2}{a} D_1} \nonumber \\
&& \phantom{ = \sum_x \bigg\{}\times \bigg( D_2 (A - i \gamma_5 B) \gamma_\mu \chi + (A + i \gamma_5 B) D_2 \gamma_\mu \chi 
 - D_2 \big[ (A - i \gamma_5 B) \gamma_\mu \chi \big] \bigg) \nonumber \\ 
&& \phantom{ = \sum_x \bigg\{} - 2 \bar \chi (A + i \gamma_5 B) D_{2 \mu} \chi -
\bar \chi (D_{2 \mu} A + i \gamma_5 D_{2 \mu} B) \chi 
\bigg\} \, ,
\label{lastchi}
\eneqa
where (\ref{delta12A}), (\ref{delta12B}) and (\ref{delta12chi}) have been used. 
Due to the Ginsparg-Wilson relation (\ref{gw1}), we have that 
\beeq 
\bigg( (1 - \frac{a}{2} D_1)^{-1} D_2 + m \bigg) \frac{m(1 - \frac{a}{2} D_1) - D_2}{m^2(1 - \frac{a}{2} D_1) + \frac{2}{a} D_1}  
= 1
\eneq
and (\ref{lastchi}) becomes
\beeqa
&&  -\frac{g}{\sqrt{2}} \sum_x \bigg\{ \bar \chi  D_2 (A - i \gamma_5 B) \gamma_\mu \chi + 
\bar \chi (A + i \gamma_5 B) D_2 \gamma_\mu \chi - \bar \chi \gamma_\mu (A - i \gamma_5 \chi) D_2 \chi \nonumber \\
&& \phantom{ = \sum_x \bigg\{} - 2 \bar \chi (A + i \gamma_5 B) D_{2 \mu} \chi - \bar \chi (D_{2 \mu} A + i \gamma_5 D_{2 \mu} B) 
\chi \bigg\}
\eneqa
where in the last term of the first line an integration by part and the relation (\ref{c}) has been used. 
Finally, terms containing the derivative of the scalar fields cancel out, while to prove the cancellation of the 
remaining ones a Fierz rearrangement is needed.

\section*{Appendix C}
By making a variation of the generating functional with respect to the lattice supersymmetry transformation 
(\ref{complete}), the Ward identity reads
\beeq
< J_\Phi \delta \Phi > = 0
\eneq
where $J_\Phi $ are the sources for the fields $\Phi = (A,B,F,G,\chi)$. A simple Ward identity is
 \beeq\label{WI}
<D_2(A - i \gamma_5 B)> + < F > - i \gamma_5 < G > - g < R > = 0 \, .
\eneq
The first term of this Ward identity is zero because of $\delta-$momentum 
conservation and $D_2(k=0)=0$. In the following, we show that this Ward identity is satisfied 
to order $O(g)$, i.e.
\beeq
< F >^{(1)} - i \gamma_5 < G >^{(1)} - g < R >^{(0)} = 0 
\eneq
where $< {\cal O} >^{(n)}$ denotes the expectation value of the function ${\cal O}$ to the $n$ order in perturbation theory.
From the action (\ref{wz2}), the free propagators are
\beeqa
< A A >^{(0)} &=& < B B >^{(0)} = -{\cal M}^{-1} (1 - \frac{a}{2} D_1)^{-1} \nonumber \\
< F F >^{(0)} &=& < G G >^{(0)} = \frac{2}{a}{\cal M}^{-1}  D_1  \nonumber \\
< A F >^{(0)} &=& < B G >^{(0)} = m \,{\cal M}^{-1} \nonumber \\
< \chi \bar \chi >^{(0)} & =& -{\cal M}^{-1} ((1 - \frac{a}{2} D_1)^{-1} D_2 - m) \, .
\eneqa
The term $< F_x >^{(1)}$  in momentum space reads (a factor $g/\sqrt{2}$ is omitted)
\beeqa
&&
<F(k)>^{(1)}=\int_{pq}\bigg\{<\bar \chi(p) \chi(q) >^{(0)} < F(k) A(-p-q) >^{(0)} 
\nonumber \\ &&
\phantom{<F(k)>^{(1)}=}
+< F(k) F(-p-q) >^{(0)} ( < A(p) A(q) >^{(0)} - < B(p) B(q) >^{(0)} )\nonumber\\&&
\phantom{<F(k)>^{(1)}=}
+ 2  < F(k) A(-p-q) >^{(0)} < F(p) A(q) >^{(0)} 
\nonumber \\ &&
\phantom{<F(k)>^{(1)}=}
+ 2  < F(k) A(-p-q) >^{(0)} < B(p) G(q) >^{(0)}\bigg\} \, .
\eneqa
By substituting the propagators we have
\beeqa
&&<F(k)>^{(1)} =  \frac{\delta^4(k)}{m} \int_{p} {\cal M}^{-1}(p) Tr((1 - \frac{a}{2} D_1(p))^{-1} D_2(p) - m)\nonumber \\
&&
\phantom{<F(k)>^{(1)} =}
 + 4  \delta^4(k) \int_{p} {\cal M}^{-1}(p) = 0.
\eneqa
The one point function of $G$ is zero at this order.

Finally, the last term of the Ward identity (\ref{WI}) is
\beeq
< R >^{(0)} = ((1 - \frac{a}{2} D_1)^{-1} D_2 + m)^{-1} <\Delta L>^{(0)} \, .
\eneq
In the following  we consider only the contribution from the  field $A$ since $B$ can be treated similarly.
In momentum space we have
\beeqa&&
< R(k)>^{(0)} = ((1 - \frac{a}{2} D_1(k))^{-1} D_2(k) + m)^{-1}\delta^4(k)\nonumber\\
&&\phantom{< R(k)>^{(0)} =}\times
\bigg\{-2\int_q D_2(q)
{\cal M}^{-1}(q) (1 - \frac{a}{2} D_1(q))^{-1}\nonumber\\
&&\phantom{< R(k)>^{(0)} =\times -}
+D_2(k)\int_q{\cal M}^{-1}(q) (1 - \frac{a}{2} D_1(q))^{-1}\bigg\}=0.
\eneqa
Indeed, the first integrand is an odd function of $q$, while the second term is zero since $D_2(k=0)=0$.

\end{document}